\documentclass[conference]{IEEEtran}

\newif\ifExtended
\Extendedtrue

\usepackage[utf8]{inputenc}

\usepackage{amsmath}
\usepackage[ruled]{algorithm2e} 

\SetAlFnt{\small}
\SetAlCapFnt{\small}
\SetAlCapNameFnt{\small}
\SetAlCapHSkip{0pt}
\IncMargin{-\parindent}

\ifCLASSOPTIONcompsoc
\usepackage[caption=false,font=normalsize,labelfont=sf,textfont=sf]{subfig}
\else
\usepackage[caption=false,font=footnotesize]{subfig}
\fi
\usepackage{dblfloatfix}
\usepackage{url}

\usepackage{enumitem}
\usepackage{verbatim}
\usepackage{amsthm}
\newtheorem{theorem}{Theorem}
\theoremstyle{definition}
\newtheorem{definition}{Definition}

\newcommand{\vect}[1]{\ensuremath{\boldsymbol{#1}}}

\newcommand{\vr}{\vect{r}}
\newcommand{\vl}{\vect{l}}
\newcommand{\vro}{\vect{ro}}
\newcommand{\vlo}{\vect{lo}}

\newcommand{\calF}[0]{\ensuremath{\mathcal{F}}}
\newcommand{\calG}[0]{\ensuremath{\mathcal{G}}}

\DeclareMathOperator{\argmin}{argmin}

\usepackage{tikz}
\usetikzlibrary{positioning,shadows,backgrounds,shapes}
\usepackage{pgfplots}
\pgfplotsset{compat=1.10}
\definecolor{PlotRed}{RGB}{228,26,28}
\definecolor{PlotBlue}{RGB}{55,126,184}
\definecolor{PlotGreen}{RGB}{77,175,74}
\definecolor{PlotOrange}{RGB}{255,127,0}

\usepackage{colortbl}
\definecolor{TableRowGray}{gray}{0.93}

\usepackage[textsize=tiny, textwidth=1.5cm, disable]{todonotes}

\newcommand{\Aron}[1]{\todo[color=yellow!45, linecolor=black!50]{\textbf{Aron}: #1}}

\begin{document}
\setlength{\marginparwidth}{1.375cm} 

\thispagestyle{plain}
\pagestyle{plain}

\title{Synergistic Security for the \\Industrial Internet of Things:\\ Integrating Redundancy, Diversity, and Hardening}

\author{\IEEEauthorblockN{Aron Laszka}
\IEEEauthorblockA{University of Houston\\
Houston, TX, USA}
\and
\IEEEauthorblockN{Waseem Abbas}
\IEEEauthorblockA{Information Technology University\\
Lahore, Pakistan}
\and
\IEEEauthorblockN{Yevgeniy Vorobeychik}
\IEEEauthorblockA{Washington University in Saint Louis\\
St. Louis, MO, USA}
\and
\IEEEauthorblockN{Xenofon Koutsoukos}
\IEEEauthorblockA{Vanderbilt University\\
Nashville, TN, USA}}

\maketitle

\ifExtended
\begin{center}
Published in the proceedings of the 2018 IEEE International Conference on Industrial Internet (ICII).
\end{center}
\fi

\begin{abstract}
As the Industrial Internet of Things (IIot) becomes more prevalent
in critical application domains, ensuring security and resilience
in the face of cyber-attacks is becoming an issue of paramount importance.
Cyber-attacks against critical infrastructures, for example, 
against smart water-distribution and transportation systems, 
pose serious threats to public health and safety.
Owing to the severity of these threats, a variety of security techniques are available.
However, no single technique can address the whole spectrum of cyber-attacks that may be launched by a determined and resourceful attacker.
In light of this, we consider a multi-pronged approach for designing secure and resilient IIoT systems, which integrates redundancy, diversity, and hardening techniques.
We introduce a framework for quantifying cyber-security risks and optimizing IIoT design by determining security investments in redundancy,
diversity, and hardening.
To demonstrate the applicability of our framework, we present \ifExtended two case studies in water distribution and transportation \fi systems.
Our numerical evaluation shows that integrating redundancy, 
diversity, and hardening can lead to reduced security risk 
at the same cost.
\end{abstract}

\section{Introduction}
\label{sec:intro}


Emerging industrial platforms such as the Industrial Internet (II) 
in the US and Industrie 4.0 in Europe are creating novel systems  
that include the devices, systems, networks, and controls used to 
operate and/or automate Industrial Internet of Things (IIoT) systems. 
IIoT systems abound in modern society, and it is not surprising 
that many of these systems are targets for attacks. Critical infrastructure
such as water management and transportation systems, in particular, have been growing more connected following recent advances in co-engineered interacting networks of physical and computational components. Due to the tightly coupled nature between the cyber and physical domains, 
new attack vectors are emerging. Attacks can include physical destruction, network spoofing, malware, data corruption, malicious insiders, and others. Further, the impacts of attacks propagate because of
tight interactions.
As IIoT systems become more ubiquitous, the risks posed by cyber-attacks 
becomes severe. 
The steady increase in the number of reported cyber-incidents evidences how difficult it is in practice to secure such systems against determined attackers. 

A variety of techniques have been proposed for providing resilience against cyber-attacks, ranging from hardening techniques (e.g., address-space layout randomization) to increasing system diversity (e.g.,~\cite{odonnell2004achieving}). 
However, defending complex and large-scale IIoT systems is particularly challenging.
These systems often face a variety of threats, have large attack surfaces, and may contain a number of undiscovered vulnerabilities. 
In light of these factors, it is clear that there is no ``silver bullet'' technique that could protect a complex system against every kind of attack.
\ifExtended

\fi
Instead of relying on a single technique, defenders must employ multi-pronged solutions, which combine multiple techniques for improving the security and resilience of IIoT.
We can divide many of existing techniques into three canonical approaches: 
\ifExtended
\begin{itemize}[topsep=0pt]
\else
\begin{itemize}[topsep=0pt, leftmargin=*]
\fi
\item \emph{Redundancy} for deploying additional redundant components in a system, so that even if some components are compromised or impaired, the system may retain normal (or at least adequate) functionality; 
\item \emph{Diversity} for implementing components using a diverse set of component types, so that vulnerabilities which are present in only a single type have limited impact on the system; and 
\item \emph{Hardening} for reinforcing individual components or component types (e.g., tamper-resistant hardware and firewalls), so that they are harder to compromise or impair.
\end{itemize}

While it is possible to combine these approaches easily by designing and implementing them 
independently, security and resilience of IIoT systems can be significantly improved by designing and implementing 
them in an integrated manner.  
However, a sound framework and methodology for combining techniques from different approaches is lacking. 
In lieu of a unified framework or methodology, defenders must follow best practices and intuition when integrating techniques, which can result in the deployment of ineffective---or even vulnerable---combinations.

In this paper, we propose a framework for integrating redundancy, diversity, and hardening techniques for designing secure and resilient
IIoT systems. The objective is to develop a systematic framework for 
prioritizing investments for reducing  security risk. The contributions of the paper are as follows:
\ifExtended
\begin{itemize}[topsep=0pt]
\else
\begin{itemize}[topsep=0pt, leftmargin=*]
\fi
\item Establishing a system model that can capture (1) a wide variety of 
components that are found in IIoT as well as the interactions between 
them, (2) a security investment model for redundancy, diversity,
and hardening, and (3) a security risk model which quantifies the impact
of attacks and defense mechanisms (Section~\ref{sec:model}).
\item Formulating the resilient IIoT design problem as an optimization 
problem for prioritizing security investments and showing that the 
problem is NP-hard (Section~\ref{sec:results}).
\item Developing an efficient meta-heuristic design algorithm \ifExtended based on 
simulated annealing \fi for finding near-optimal designs in practice  (Section~\ref{sec:results}).
\item Evaluating the applicability of the approach using two case 
studies in canonical IIoT  domains of water distribution and 
transportation systems (Sections~\ref{sec:application} and~\ref{sec:numerical}).
\end{itemize}
We give an overview of related work in Section~\ref{sec:related}\ifExtended ~and provide concluding remarks in Section~\ref{sec:concl}\fi.


\section{Model}
\label{sec:model}

\ifExtended
\newcommand{\exampleCPSscaleX}{0.99cm}
\newcommand{\exampleCPSscaleY}{1.21cm}
\fi

\ifExtended
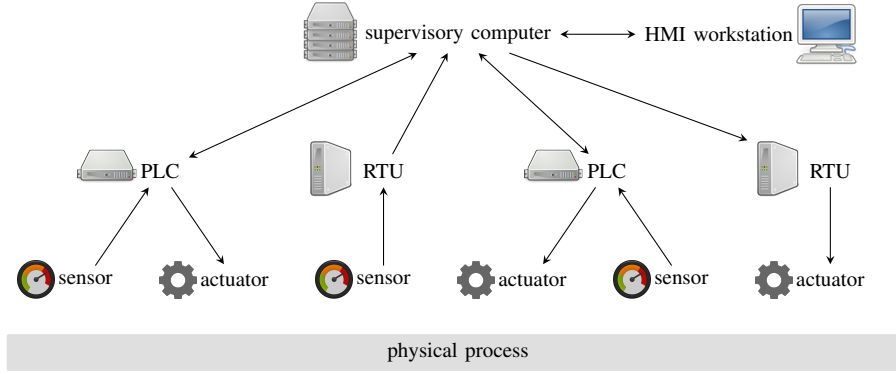
\begin{figure*}
\else
\begin{figure}
\fi
\centering
\begin{tikzpicture}[x=\exampleCPSscaleX, y=\exampleCPSscaleY, font=\footnotesize,
  link/.style={->, >=stealth},
  component/.style={fill=white, draw=black, minimum size=0.5cm, rounded corners=0.025cm},
  supervisor/.style={label={[label distance=-0.2cm]left:\includegraphics[width=0.8cm]{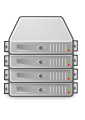}}},
  PLC/.style={label={[label distance=-0.2cm]left:\includegraphics[width=0.8cm]{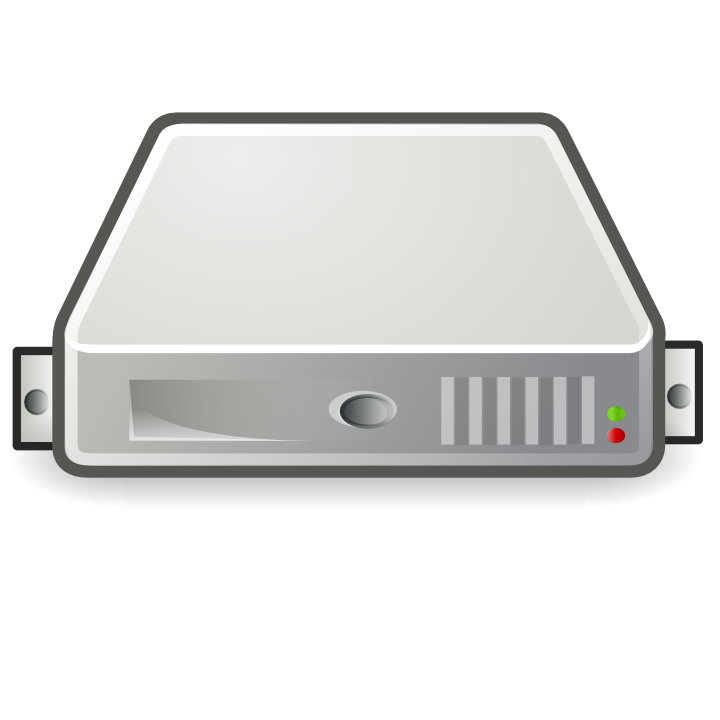}}},
  RTU/.style={label={[label distance=-0.2cm]left:\includegraphics[width=0.8cm]{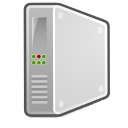}}},
  sensor/.style={label={[label distance=-0.2cm]left:\includegraphics[width=0.5cm]{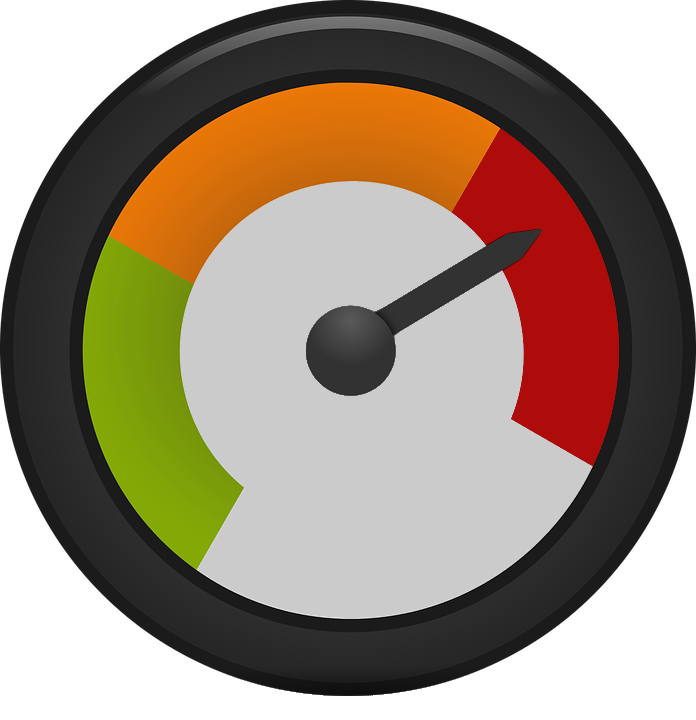}}},
  actuator/.style={label={[label distance=-0.2cm]left:\includegraphics[width=0.5cm]{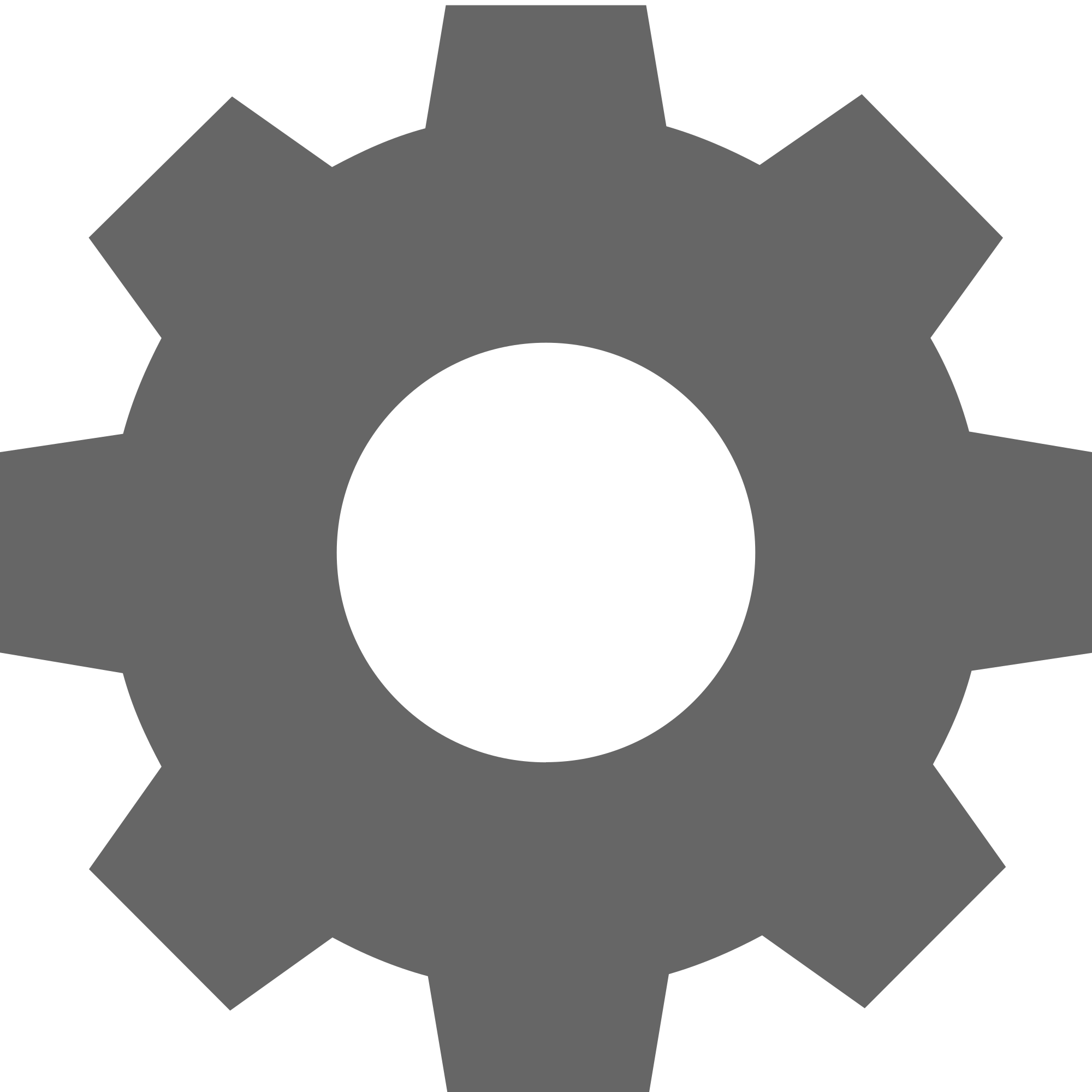}}},
  HMI/.style={label={[label distance=-0.2cm]right:\includegraphics[width=0.8cm]{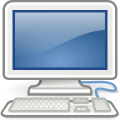}}}, 
]
\ifExtended
\node [fill=black!12.5, rounded corners=0.025cm, minimum width=12cm] (physical) at (0, 0) {physical process};
\fi
\foreach \pos/\name in {{-5/sensor1}, {-1/sensor2}, {3/sensor3}}
  \node [sensor] (\name) at (\pos, 0.8) {sensor};
\foreach \pos/\name in {{-3/actuator1}, {1/actuator2}, {5/actuator3}}
  \node [actuator] (\name) at (\pos, 0.8) {actuator};
\foreach \pos/\name in {{-4/plc1}, {2/plc2}}
  \node [PLC] (\name) at (\pos, 2) {PLC};
\foreach \pos/\name in {{-1/rtu1}, {5/rtu2}}
  \node [RTU] (\name) at (\pos, 2) {RTU};
\node [supervisor] (supervisor1) at (0, 3.5) {supervisory computer};
\ifExtended
\node [HMI, right=1cm of supervisor1] (hmi1) {HMI workstation};
\fi
\begin{scope}[on background layer]
\draw [link] (sensor1) -- (plc1);
\draw [link] (plc1) -- (actuator1);
\draw [link] (plc2) -- (actuator2);
\draw [link] (sensor3) -- (plc2);
\draw [link] (sensor2) -- (rtu1);
\draw [link] (rtu2) -- (actuator3);
\draw [link, <->] (plc1) -- (supervisor1);
\draw [link] (rtu1) -- (supervisor1);
\draw [link, <->] (plc2) -- (supervisor1);
\draw [link, shorten >=0.75cm] (supervisor1) -- (rtu2);
\draw [link, <->] (supervisor1) -- (hmi1);
\end{scope}
\end{tikzpicture}
\caption{Example cyber-physical system. Arrows represent flows of sensor data and control signals.}
\label{fig:exampleCPS}
\ifExtended
\end{figure*}
\else
\end{figure}
\fi

An IIoT system is comprised of a variety of components: sensors, controllers, actuators, and human-machine interfaces for interacting 
with users as shown in Figure~\ref{fig:exampleCPS}.
Our first step introduces a general system model for 
evaluating security risk. 
First, we present a high-level model of IIoT systemts. Then, we introduce a model of security investments in redundancy, diversity, and hardening,
and we quantify risks posed by cyber-attacks, considering both probability and impact.
Based on this model, we formulate the problem of optimal system design.
For a list of symbols used in this paper, see \ifExtended Table~\ref{tab:symbols}\fi.

\ifExtended
\begin{table}[h]
\caption{List of Symbols}
\vspace{-0.6em}
\centering
\label{tab:symbols}
\renewcommand*{\arraystretch}{1.3}
\begin{tabular}{|l|p{6.9cm}|}
\hline
Symbol & Description \\
\hline
\multicolumn{2}{|c|}{Constants} \\
\hline
$C$ & set of components \\
\rowcolor{TableRowGray} $E$ & set of connections between components \\
$O_c$ & set of components connecting to component $c \in C$ \\
\rowcolor{TableRowGray} $T_c$ & type of component $c \in C$ \\
$I$ & set of implementation types \\
\rowcolor{TableRowGray} $I_c$ & set of implementation types available for component $c \in C$ \\
$R_i$ & cost of deploying an instance of implementation type $i \in I$ \\
\rowcolor{TableRowGray} $D_i$ & cost of deploying at least of instance of type $i \in I$ \\
$L_i$ & set of hardening levels available for type $i \in I$ \\
\rowcolor{TableRowGray} $S_l$ & probability that hardening level $l \in L_i$ is secure \\
$H_l$ & cost of attaining hardening level $l \in L_i$ \\
\hline
\multicolumn{2}{|c|}{Deployment} \\
\hline
$r_c$ & set of implementation types deployed for component $c \in C$ \\
\rowcolor{TableRowGray} $l_i$ & hardening level chosen for implementation type $i \in I$ \\
\hline
\end{tabular}
\end{table}
\fi

\subsection{System Model}
\label{sec:systemModel}

We model the cyber part of the system as a directed graph $G = (C, E)$.
The set of nodes $C$ represents the components of the system, while the set of directed edges $E$ represents connections between the components, which are used to send data and control signals.
For each component $c \in C$, we let $O_c \subseteq C$ denote the set of origin components of the incoming edges of component $c$. 
Further, we let $T_c$ denote the type of component $c$, which is one of the following:
\begin{itemize}[topsep=0pt]
\item $\textit{sensor}$: components that measure the state of physical processes (e.g., pressure sensors); 
\item $\textit{actuator}$: components that directly affect physical processes (e.g., valves); 
\item $\textit{processing}$: components that process and store data and control signals (e.g., PLCs); 
\item $\textit{interface}$: components that interact with human users (e.g., HMI workstations).
\end{itemize}

The implementation of each component is chosen from a set of implementation types.
We let $I_c$ denote the set of types that may be used to implement component $c$, and we let $I$ denote the set of all implementation types that may be used in the system (i.e., $I = \cup_{c \in C} I_c$).

\subsection{Security Investment Model}
\label{sec:securityInvestment}


\subsubsection{Redundancy}

We model redundancy as deploying multiple instances of the same component.
For simplicity, we assume that for each component, at most one instance of each suitable implementation type is deployed.
\footnote{Note that relaxing this assumption would be straightforward; however, such a generalization would provide little further insight into security.}
We make this assumption because our goal is to address security risks posed by deliberate attacks, and if a security vulnerability exists in an implementation type, then attackers can typically compromise all instances of that type.

We let $r_c \subseteq I_c$ denote the set of implementation types that are deployed for component $c \in C$. 
To quantify the cost of redundancy, we let $R_i$ denote the cost of deploying an instance of type $i \in I_c$.
Then, the total cost of redundancy is
\begin{equation}
\text{cost of redundancy } = \sum_{c \in C} \sum_{i \in r_c} R_i .
\end{equation}

\subsubsection{Diversity}

We model diversity as deploying a diverse set of implementation types.
In other words, diversity is modeled as selecting different implementations $r_c$ to be deployed for each component $c \in C$ (or at least attempting to use as many distinct sets as possible).

To quantify the cost of diversity, we let $D_i$ denote the cost of using an implementation type $i \in I$ in any non-zero number of components (i.e., $D_i$ is the cost incurred when the first instance of type $i$ is deployed).
Then, the cost of diversity is
\begin{equation}
\text{cost of diversity } = \sum_{i \,\in\, \bigcup_{c \in C} r_c} D_i .
\end{equation}

\subsubsection{Hardening}

We model the hardening of an implementation type as decreasing the probability that a zero-day security vulnerability is discovered by an attacker.
We assume that hardening is applied in steps (e.g., performing a code review), resulting in a discrete set of hardening levels.

We let $L_i$ denote the set of hardening levels available for implementation type $i \in I$, and we let $l_i$ denote the chosen level.
To model the amount of security provided by hardening level $l \in L_i$, we let $S_l$ denote the probability that the implementation type will be secure (i.e, no zero-day vulnerability is discovered) if level $l$ is chosen.
To quantify the cost of hardening, we let~$H_l$ denote the cost of attaining level $l \in L_i$.
Then, the total cost of hardening is
\begin{equation}
\text{cost of hardening } = \sum_{i \in I} H_{l_i} .
\end{equation}

\subsection{Security Risk Model}
\label{sec:securityRisk}

Next, we quantify the risks faced by a system with given redundancy, diversity, and hardening design.
In principle, risk can be quantified as
\begin{equation}
\textit{Risk} = \sum_{\textit{outcome}} \Pr[\textit{outcome}] \cdot \textit{Impact}(\textit{outcome}) .
\end{equation}

In our model, an outcome can be represented as a set of components that have been compromised by an attacker:
\begin{equation}
\textit{Risk}(\vr, \vl) = \sum_{\hat{C} \subseteq C} \Pr[\hat{C} \text{ is compromised}] \cdot \textit{Impact}(\hat{C}) ,
\end{equation}
where $\textit{Impact}(\hat{C})$ is the amount of loss inflicted on the system by an attacker who has compromised components $\hat{C}$.
In the remainder of this subsection, we discuss how to measure $\Pr[\hat{C} \text{ is compromised}]$ and $\textit{Impact}(\hat{C})$.

\subsubsection{Probability}

We quantify the probability that an attacker compromises a set of components $\hat{C} \subseteq C$ \emph{implicitly} by describing a probabilistic process that models how an attacker can take control of the components of a system one-by-one.
\Aron{We could also consider a combination of the two models (e.g., consider the sum risks from both stealthy and non-stealthy attacks).}
We consider two alternative attack models in our framework: non-stealthy attacks and stealthy attacks.
The two attack models are summarized in Table~\ref{tab:comprRules}.

\begin{table}[h]
\begin{center}
\caption{Component Compromise Rules}
\label{tab:comprRules}
\vspace{-0.5em}
\renewcommand*{\arraystretch}{1.4}
\begin{tabular}{|p{1cm}|p{1.95cm}lll|}
\hline
Attack & \multicolumn{4}{c|}{Component Type} \\
\cline{2-5}
Type & sensor & actuator & processing & interface \\
\hline
stealthy attack & if all instances are compromised & \multicolumn{3}{p{4.3cm}|}{if all instances are compromised or 
all input components are compromised} \\
\hline
non-stealthy attack & if majority of instances are compromised & \multicolumn{3}{p{4.3cm}|}{if majority of instances are compromised or majority of input components are compromised} \\
\hline
\end{tabular}
\end{center}
\end{table}

\paragraph{Non-Stealthy Attacks}
\label{sec:nonstealthy}
First, an attacker attempts to find exploitable vulnerabilities in the implementation types that are deployed in the system.
Based on our hardening model, the attacker discovers a zero-day vulnerability in each implementation type $i \in I$ with probability $1 - S_{l_i}$ (independently of the other types).
We then consider all instances of the vulnerable implementation types to be compromised, and let $\hat{I}$ denote the set of vulnerable implementations.

Next, we determine the set of compromised components $\hat{C}$.
We start with $\hat{C} = \emptyset$, and then extend the set $\hat{C}$ in iterations based on the following rules:
\begin{itemize}[topsep=0pt]
\item a \emph{sensor} component $c$ is considered to be compromised if the majority of its instances $r_c$ are vulnerable (i.e., if $|r_c \cap \hat{I}| \geq |r_c| / 2$), 
\item an \emph{actuator}, \emph{processing}, or \emph{interface} component $c$ is considered to be compromised if the majority of its instances $r_c$ are vulnerable or if the majority of its inputs are compromised (i.e., if $|O_c \cap \hat{C}| \geq |O_c| / 2$). 
\end{itemize}
We repeat the above steps until the set of compromised components $\hat{C}$ cannot be extended any further.

\paragraph{Stealthy Attacks}
\label{sec:stealthy}
For stealthy attacks, the process is the same except that ``majority'' is replaced in both rules with ``all'' (i.e., $|r_c \cap \hat{I}| = |r_c|$ and $|O_c \cap \hat{C}| = |O_c|$).

\subsubsection{Impact}

We let $\textit{Impact}(\hat{C})$ denote the financial and physical loss resulting from an attack that compromises and maliciously controls components in $\hat{C}$.
The exact formulation of $\textit{Impact}(\hat{C})$ depends on the system and the characteristics of its physical processes.
\ifExtended
In this paper, we consider two types of systems, water-distribution and transportation systems, which we will describe in detail in Section~\ref{sec:application}.
\fi

\ifExtended
\subsection{Optimal Design Problem}
\label{sec:problemFormulation}
%

We first formulate the problem with fixed investments in redundancy, diversity, and hardening.
\begin{definition}[Optimal Design Problem (Fixed Redundancy, Diversity, and Hardening)]
\else
Finally, we formulate the problem of finding an optimal design as follows.
\begin{definition}[Optimal Design Problem]
\fi
Given redundancy, diversity, and hardening investments $R$, $D$, and $H$,  an \emph{optimal design} $(\vr, \vl)$~is
\begin{equation}
\argmin_{\vr, \vl} \textit{Risk}(\vr, \vl)
\end{equation}
subject to
\ifExtended
\begin{align}
\forall c \in C: ~ r_c &\subseteq I_c \\
\forall l \in I: ~ l_i &\in L_i \\
\sum_{c \in C} \sum_{i \in r_c} R_i &\leq R \\
\sum_{i \in \cup_{c \in C} r_c} D_i &\leq D \\
\sum_{i \in I} H_{l_i} &\leq H .
\end{align}
\fi
\end{definition}

\ifExtended
Next, we introduce a more general formulation, in which we can determine the amounts to invest in redundancy, diversity, and hardening.

\begin{definition}[Optimal Design Problem]
Given security budget~$B$, an \emph{optimal design} $(\vr, \vl)$ is
\begin{equation}
\argmin_{\vr, \vl} \textit{Risk}(\vr, \vl)
\end{equation}
subject to
\begin{align}
\forall c \in C: ~ r_c &\subseteq I_c \label{eq:optResDesProbConstr1} \\
\forall l \in I: ~ l_i &\in L_i \label{eq:optResDesProbConstr2} \\
\sum_{c \in C} \sum_{i \in r_c} R_i + \sum_{i \in \cup_{c \in C} r_c} D_i + \sum_{i \in I} H_{l_i} &\leq B . \label{eq:optResDesProbConstr3}
\end{align}
\end{definition}
\fi

\section{Computational Analysis and Meta-Heuristic Algorithms}
\label{sec:results}

Since the number of feasible designs to choose from may be very large even for small systems, finding an optimal design using exhaustive search is computationally infeasible.
In light of this, a key question for the practical application of the proposed framework is whether there exist efficient algorithms for finding optimal or near-optimal designs.
We first show that finding an optimal design is computationally \ifExtended challenging by showing that the problem is NP-hard. \fi
Then, we introduce an efficient meta-heuristic algorithm that can find a near-optimal solution in polynomial time.

\subsection{Computational Complexity}

The objective of the design problem depends on the impact function, which could be any function, even one that is hard to compute.
To show that the design problem is inherently hard (not only due to the potential complexity of computing the impact function), we \ifExtended consider computational complexity assuming \fi a simplistic impact function, whose value is simply the number of compromised components.
Formally, we consider $\textit{Impact}(\hat{C}) = |\hat{C}|$.

\ifExtended
To show that the optimal design problem is NP-hard, we first introduce a decision version of the problem.

\begin{definition}[Optimal Design Problem (Decision Version)]
Given security budget $B$ and threshold risk $\textit{Risk}^*$, determine if there exists a design $(\vr, \vl)$ such that $\textit{Risk}(\vr, \vl) \leq \textit{Risk}^*$ and Equations \eqref{eq:optResDesProbConstr1}, \eqref{eq:optResDesProbConstr2}, and \eqref{eq:optResDesProbConstr3} hold.
\end{definition}

We will show that the above problem is NP-hard using a reduction from a well-known NP-hard problem, the Set Cover Problem, which is defined as follows.

\begin{definition}[Set Cover Problem]
Given a set $U$, a set $\calF$ of subsets of $U$, and a threshold $k$, find a subset $\calG \subseteq \calF$ consisting of at most $k$ subsets such that $\calG$ covers $U$ (i.e., for every $u \in U$, there exists a $g \in \calG$ such that $u \in g$).
\end{definition}
\fi

\begin{theorem}
\label{thm:NPhard}
The Optimal Design Problem is NP-hard.
\end{theorem}

\ifExtended
\begin{IEEEproof}
Given an instance $(U, \calF, k)$ of the Set Cover Problem (SCP), we construct an instance of the Optimal Design Problem (ODP) with stealthy attacks as follows:
\begin{itemize}[topsep=0pt]
\item let $C := U$, $E := \emptyset$, and $I := \calF$,
\item for every $c \in C$, let $T_c := $ \textit{sensor},
\item for every $c \in C$, let $I_c := \{ i \in \calF \,|\, c \in i \}$,
\item for every $i \in I$, let $R_i := 0$,
\item for every $i \in I$, let $D_i := 0$,
\item for every $i \in I$, let $L_i := \{\textit{insecure}, \textit{secure}\}$,
\item let $H_{\textit{insecure}} := 0$ and $S_{\textit{insecure}} := 0$,
\item let $H_{\textit{secure}} := 1$ and $S_{\textit{secure}} := 1$, 
\item let $B := k$ and $\textit{Risk}^* := 0$.
\end{itemize}
Clearly, the above reduction can be performed in a polynomial number of steps.
It remains to show that the constructed instance of the ODP has a solution if and only if the SCP instance has a solution.

First, suppose that the SCP instance has a solution $\calG$.
Then, we show that there exists feasible design $(\vr, \vl)$ that is a solution to the ODP instance.
For every component $c \in C$, let $r_c = I_c$.
For every implementation type $i \in I$, let $l_i = \textit{secure}$ if $i \in \calG$ (recall that in the construction of the ODP instance, we let the implementation types $I$ correspond to the set of subsets $\calF$, and the solution $\calG$ is a subset of $\calF$) and let $l_i = \textit{insecure}$ if $i \not\in \calG$.
Clearly, this is a feasible design since its hardening cost is 
\begin{align}
\sum_{i \in I} H_{l_i} &= \sum_{i \in \calG} H_{\textit{secure}} \sum_{i \in I \setminus \calG} H_{\textit{insecure}} \\
&= \sum_{i \in \calG} 1 \sum_{i \in I \setminus \calG} 0 \\
&= |\calG| \leq k = B ,
\end{align}
and all other costs are zero.
Since $S_{\textit{secure}} = 0$, implementation types from $\calG$ are never vulnerable, and any component $c$ that has at least one secure implementation type (i.e., $I_c \cap \calG \neq 0$) is never compromised by a stealthy attack.
If $\calG$ is a set cover, then there exists at least one secure implementation type $i \in \calG$ for each $c$ such that $i \in I_c$, which implies that no component will be compromised.
Therefore, $\hat{C} = \emptyset$ is the only possible outcome, which implies that $\textit{Risk}(\vr, \vl) = 0$ as $\textit{Impact}(\emptyset) = 0$ by definition.

Second, suppose that the ODP instance has a solution $(\vr, \vl)$.
Then, we can show that there exists a solution $\calG$ to the SCP instance.
Let $\calG = \left\{ i \in \calF \,\middle|\, l_i = \textit{secure} \right\}$ (i.e., the set of implementation types that are secure).
Clearly, $\calG$ is a feasible solution due to the budget constraint.
Next, using an argument that is similar to the one that we used in the previous case, we can show that if $\calG$ was not a set cover, then $\textit{Risk}(\vr, \vl)$ would be greater than zero.
The claim of the theorem then follows from this readily.
\end{IEEEproof}
\fi

\subsection{Meta-Heuristic Design Algorithm}

We propose an efficient meta-heuristic algorithm for finding near-optimal designs in practice.
Our algorithm is based on simulated annealing, which requires randomly generating feasible solutions that are ``neighbors'' of (i.e., similar to) a given solution. 
Unfortunately, in our solution space (i.e., in the set of designs that satisfy the budget constraints), the feasible neighbors of a solution are not naturally defined.
Hence, \ifExtended before we present our meta-heuristic algorithm, \fi we first introduce an alternative representation of feasible designs, which we call design plans.

\begin{definition}[Design Plan]
A \emph{design plan} is a pair $(\vro, \vlo)$, where
\begin{itemize}[topsep=0pt] 
\item $\vro$ is a list of component-implementation pairs~$(c, i) \in C \times I$ such that $i \in I_c$ holds for every pair $(c, i) \in \vro$, and each possible pair $(c, i)$ appears exactly once in $\vro$;
\item $\vlo$ is an ordered multiset of implementation types such that each implementation type $i \in I$ appears exactly $|L_i| - 1$ times in $\vlo$.
\end{itemize}
\end{definition}

\begin{algorithm}[h!]
\caption{\textit{MapToDesign}$(\vro, \vlo)$}
\label{alg:des_rep}
\KwData{optimal design problem, list $\vro$, ordered multiset $\vlo$} 
\KwResult{design $(\vr, \vl)$}
$\forall c \in C: ~ r_c \gets \emptyset$\ifExtended \\ \fi
$\forall i \in I: ~ l_i \gets \argmin_{l \in L_i} H_l$ \\
\For{$(c, i) \in \vro$}{
  $\vr' \gets \vr$ \ifExtended \\ \fi
  ${r'}_c \gets r_c \cup \{i\}$ \\
  \If{$(\vr', \vl)$ is feasible} {
    $\vr \gets \vr'$
  }
}
\For{$i \in \vlo$}{
  $\vl' \gets \vl$\ifExtended \\ \fi
  ${l'}_i \gets \argmin_{l \in L_i : H_l > H_{l_i}} H_l$ \\
  \If{$(\vr, \vl')$ is feasible} {
    $\vl \gets \vl'$
  }
}
\textbf{output} $(\vr, \vl)$ 
\end{algorithm}

Next, we show how to translate a design plan $(\vro, \vlo)$ into a feasible design.
The translation is presented formally in Algorithm~\ref{alg:des_rep}\ifExtended. \fi
\ifExtended
Given redundancy, diversity, and hardening investments~$R$, $D$, and $H$, we can obtain a feasible design~$(\vr, \vl)$ as follows:
start from an empty design (i.e., no implementations deployed and lowest-cost hardening level chosen for every implementation type);
iterate over $\vro$ in order and for each $(c, i) \in \vro$, add $i$ to $r_c$ if it does not lead to the violation of the budget constraints;
finally, iterate over $\vlo$ in order and for each $i \in \vlo$, increase security level $l_i$ if it does not lead to the violation of the budget constraint.
\fi
Note this mapping is surjective.

\begin{algorithm}[h!]
\caption{Meta-Heuristic Design Algorithm}
\label{alg:sim_ann}
\KwData{optimal design problem, number of iterations $k_{\max}$, initial temperature $T_0$, cooling parameter $\beta$} 
\KwResult{design $(\vr, \vl)$}
choose $(\vro, \vlo)$ at random \\
$\rho \gets \textit{Risk}(\textit{MapToDesign}(\vro, \vlo))$ \\
\For{$k = 1, \ldots, k_{\max}$}{
  $(\vro', \vlo') \gets \textit{Perturb}(\vro, \vlo)$ \\
  $\rho' \gets \textit{Risk}(\textit{MapToDesign}(\vro', \vlo'))$ \\
  $T \gets T_0 \cdot e^{-\beta k}$ \ifExtended \\ \fi
  $pr \gets e^{(\rho' - \rho) / T}$ \\
  \If{$(\rho' < \rho) \; \vee \; (\mathtt{rand}(0, 1) \le pr)$}{
    $\vro \gets \vro'$\ifExtended \\ \fi
    $\vlo \gets \vlo'$ 
  }
}
\textbf{output} $\textit{MapToDesign}(\vro, \vlo)$ 
\end{algorithm}

Finally, we present our meta-heuristic design algorithm (see Algorithm~\ref{alg:sim_ann}), which can find a near-optimal design in polynomial time.
The algorithm starts by choosing a random design plan $(\vro, \vlo)$.
In practice, we can implement this simply as choosing a random permutation of the list of component-implementation pairs and a random permutation of the multiset of implementation types. 
\ifExtended
The algorithm then performs a fixed number of iterations, in each iteration choosing a random neighbor $(\vro', \vlo')$ of the current plan $(\vro, \vlo)$, and replacing the current plan with the neighbor with some probability.
This probability depends on the risk of both the current and the neighboring plan, and decreases with the number of iterations, as we approach the final solution.
A key step of the algorithm is $\textit{Perturb}(\vro, \vlo)$, which chooses a random neighbor of $(\vro, \vlo)$.
In practice, we implement this as taking two elements of $\vro$ at random and switching them with each other, by similarly switching the order of two random elements of $\vlo$, and returning the re-ordered list and multiset as the neighbor $(\vro, \vlo)$.
\fi

\section{Evaluation}
\label{sec:application}

To demonstrate the applicability of our framework, 
\ifExtended
we present two case studies from two canonical IIoT domains: water distribution and transportation systems.
\fi

\ifExtended
\subsection{Cyber-Physical Contamination Attacks Against Water-Distribution Networks}
\fi

IIoT systems have a particularly significant and wide application in  water distribution systems. Examples include monitoring water 
quality and detecting leaks.
On the one hand, IIoT offers significant advantages, such as
improved service and better maintenance at a low cost, but on the
other hand, potential challenges include cost of the cyber 
infrastructure, reliability of communications, and of course,
cyber-security.

As evidenced by the recent water crisis in Flint, MI~\cite{kennedy2016lead}, ensuring the quality of drinking water is of critical importance.
Compromising systems that control the treatment and distribution of drinking water may allow adversaries to suppress warnings about contaminations or to decrease the quality of water~\cite{taormina2017characterizing}.
Cyber-attacks can also have a devastating environmental impact.
For example, in 2000, a disgruntled ex-employee launched a series of attacks against the SCADA system controlling sewage equipment in Maroochy Shire, Australia~\cite{abrams2008malicious,slay2007lessons}.
As a result of these attacks, approximately 800,000 liters of raw sewage spilt out into local parks and rivers, killing marine life. 

Here, we apply our framework to model cyber-physical contamination attacks against water-distribution systems.
The system is modeled as a graph, in which links represent pipes, and nodes represent junctions of pipes, residential consumers, reservoirs, pumps, etc.
IIoT components include:
\begin{itemize}[topsep=0pt]
\item \textit{Sensors}: water-quality sensors, which are located at certain nodes of the water-distribution network;
\item \textit{Processing}: components that collect, process, and forward water-quality data;
\item \textit{Interfaces}: components with human-machine interfaces, which can alert operators about contaminations.
\end{itemize}
We consider a malicious adversary who tries to cause harm by contaminating the water network with harmful chemicals.
We assume that the adversary can introduce contaminants at certain nodes, such as unprotected reservoirs or tanks, which will then spread in the network, eventually reaching the residential consumers.
We measure the impact of this physical attack as the amount of contaminants consumed by residential consumers before the detection of the attack. 

To detect contaminations, each sensor continuously monitors the water flowing through the node at which it is deployed, and raises an alarm when the concentration of a contaminant reaches a threshold level.
The alert generated by a \emph{sensor} node is sent to a \emph{processing} node, which forwards the alert to an \emph{interface} node that can notify operators.
Once operators are alerted, they respond immediately by warning residents not to consume water from the network.

We measure the impact of a physical attack as the amount of contaminants consumed by residential consumers before they are warned.
This amount depends on the time between the physical attack and its detection, the contaminant concentration levels at the consumer nodes in this time interval, and the amount of water consumed in this interval.
\ifExtended
Note that this impact depends on the uncompromised components $C \setminus \hat{C}$ since the time of detection depends on the functionality of these components.

\fi
To increase the impact of the physical attack, the adversary launches a cyber-attack, which compromises and disables some of the components $\hat{C}$.
Since the adversary's goal is to suppress warnings, this attack can be modeled as a \emph{stealthy attack} (Section~\ref{sec:stealthy}).
We assume that the adversary first compromises a set of components $\hat{C}$, and then decides where to introduce the contaminant, maximizing the impact $\textit{Impact}(\hat{C})$. 
\ifExtended
Our goal is to minimize the risks posed by such cyber-physical attacks by designing a resilient system based on a systematic allocation of investments to redundancy, diversity, and hardening. 
We present numerical results for this case study in Section \ref{sec:numerical}.
\fi

\ifExtended

\subsection{Cyber-Attacks Against Transportation Networks}

Transportation systems is another application domain that can benefit
greatly from IIoT by driving down costs and minimizing system failures, 
while supplying vast amounts of data for operators, drivers, and facilities that result in significant operational improvements. 
Transportation systems include multiple components that are becoming susceptible to attacks through wireless interfaces or even remote attacks through the Internet~\cite{laszka2016vulnerability}.
Indeed, recent studies have shown that many traffic lights deployed in practice have easily exploitable vulnerabilities, which could allow an attacker to tamper with the configuration of these devices.
Due to hardware-based failsafes, compromising a traffic signal does not allow an attacker to set the signal into an unsafe configuration that could immediately lead to traffic accidents~\cite{ghena2014green}.
However, compromising a signal does enable tampering with its schedule, which allows an attacker to cause disastrous traffic congestions.

Here, we apply the proposed framework to model cyber-attacks against traffic control.
The physical part of the system may be modeled using a traffic model, such as Daganzo's well-known cell-transmission model~\cite{daganzo1994cell}.
The cyber-part of the system is compromised of the following components:
\begin{itemize}[topsep=0pt]
\item \emph{Interface}: components with human-machine interfaces, which operators use to control the schedules of traffic lights in the transportation network;
\item \emph{Processing}: components that process and forward control signals sent by operators;
\item \emph{Actuator}: traffic lights with software based controllers.
\end{itemize}
We consider a malicious adversary who tries to cause damage by compromising some components $\hat{C}$ of the traffic-control system and tampering with the schedules of traffic lights.
We measure the impact $\textit{Impact}(\hat{C})$ of this cyber-attack as the increase in traffic congestion, which is quantified as the total travel time of the vehicles in the network, compared to normal congestion without an attack.
We assume that the adversary aims to cause maximum damage without attempting to hide its attack.
Hence, we model its attack as a \emph{non-stealthy attack} (Section~\ref{sec:nonstealthy}). 

\fi

\section{Numerical Results}
\label{sec:numerical}

\ifExtended
In this section, we present numerical results to evaluate the proposed approach.
First, we focus on the evaluation of the approach in terms of reducing
the security risks by integrating redundancy, diversity, and hardening.
Then, we focus on the performance of the proposed design algorithm in terms of running time.
\fi

\ifExtended
\subsection{Case-Study Examples}
\fi


\ifExtended
\subsubsection{Water Distribution System}
\fi

We use a real-world water-distribution network from Kentucky, which we obtained from the Water Distribution System Research Database\ifExtended~\footnote{http://www.uky.edu/WDST/database.html}\fi\cite{jolly2014research}.
The topology of this network, which is called KY3 in the database, is shown by Figure~\ref{fig:ky3_attack_R-1_hour_2}.
In addition to  topology, the database also contains hourly water-demand values for each network node. 

\begin{figure}[h!]
\includegraphics[width=\columnwidth]{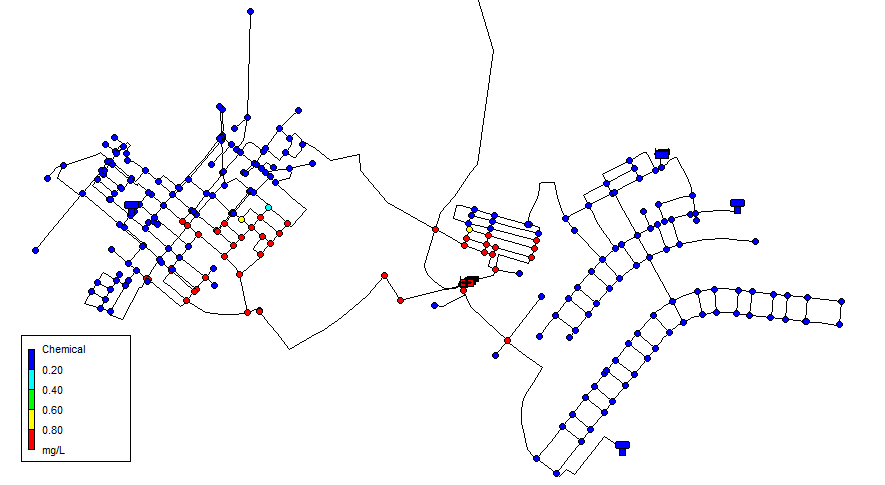}
\caption{Topology of the water-distribution network.
Colors show the spread of the contaminant from the first reservoir two hours after its introduction.}
\label{fig:ky3_attack_R-1_hour_2}
\end{figure}

We assume that the adversary can introduce a contaminant at one of six given nodes in the network, which model three tanks and three reservoirs.
Once the contaminant is introduced, we simulate its spread throughout the network using EPANET\ifExtended~\footnote{\url{https://www.epa.gov/water-research/epanet}}\fi.
From the simulation, we obtain the contaminant concentration values 
at the various nodes as functions of time.
For a given set of compromised components $\hat{C}$, we then use these values to compute the time of detection and the resulting impact $\textit{Impact}(\hat{C})$ (i.e., amount of contaminant consumed by the time of detection). 
Finally, we use the following numerical parameter values:
\ifExtended \begin{itemize}[topsep=0pt] \fi
\ifExtended \item \fi $I = \{i_1, i_2, i_3, i_4, i_5 \}$;
\ifExtended \item \fi for every $c \in C$, $I_c = I$;
\ifExtended \item \fi $R_{i_1} = R_{i_2} = R_{i_3} = 0$\ifExtended\footnote{We set these to zero to model existing deployment since we are interested in how to invest in improving security and resilience.}\fi and $R_{i_4} = R_{i_5} = 1$;
\ifExtended \item \fi $D_{i_1} = 0$\ifExtended\footnote{We set this to zero so that there always exists a feasible deployment.}\fi and $D_i = 1$ for every $i \in \{ i_2, i_3, i_4, i_5 \}$;
\ifExtended \item \fi for every $i \in I$, $L_i = \{ 1, 2, 3, \ldots, 10 \}$;
\ifExtended \item \fi for every $l \in L_i$, $S_l = 1 - 0.5^{0.5 \cdot l + 1}$ and $H_l = 4 \cdot l^2$.
\ifExtended \end{itemize} \fi

\ifExtended

\subsubsection{Transportation Network}

We use the Grid model with Random Edges (GRE) to generate a random network topology~\cite{peng2012random}, which closely resembles real-world transportation networks.\footnote{We instantiated the model with $W = 5$, $L = 5$, $p = 0.507$, and $q = 0.2761$ based on~\cite{peng2012random}.}
For a detailed description of this model, we refer the reader to~\cite{peng2012random,peng2014random}.
We use Daganzo's cell transmission model to simulate traffic flowing through the generated network~\cite{daganzo1994cell}, computing the turn decisions of the vehicles based on a linear program that minimizes total travel time~\cite{ziliaskopoulos2000linear}.
Following Daganzo's proposition, we model traffic lights as constraints on the inflow proportions~\cite{daganzo1995cell}, and we select the default (i.e., uncompromised) schedules of the traffic lights to minimize congestion.
Finally, we allow the attacker to select any valid configuration for compromised lights.

We use the following parameter values for our illustrations:
\begin{itemize}[topsep=0pt]
\item $I = \{i_1, i_2, i_3, i_4, i_5 \}$;
\item for every $c \in C$, $I_c = I$;
\item $D_{i_1} = 0$ and $D_i = 20$ for every $i \in \{ i_2, i_3, i_4, i_5 \}$;
\item for every $i \in I$, $R_i = 1$, $D_i = 20$, and $L_i = \{ 1, 2, 3, \ldots, 10 \}$;
\item for every $l \in L_i$, $S_l = 1 - 0.5^{0.5 \cdot l + 2}$ and $H_l = 10 \cdot l^2$.
\end{itemize}

\fi

\ifExtended
\subsection{Risk Evaluation}
\fi

\ifExtended
\newcommand{\figureHeight}{0.7\columnwidth}
\fi

\ifExtended
Next, we study how security risks depend on investments into redundancy, diversity, and hardening, as well as their optimal combinations.
\fi

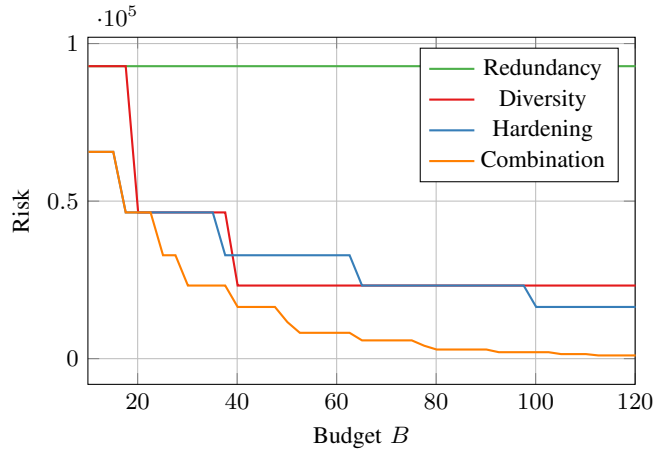
\begin{figure}[h!]
\begin{tikzpicture}[semithick]
\begin{axis}[
  width=\columnwidth,
  height=\figureHeight,
  legend pos=north east,
  xlabel=Budget $B$,
  ylabel=Risk,
  grid=major,
  font=\small,
  xmin=10,
  xmax=120
]
\addplot[no markers, solid, PlotGreen, thick] table[x=budget, y=redundancy, comment chars={\%}, col sep=comma] {data/waterRisk.csv};
\addlegendentry{Redundancy};
\addplot[no markers, solid, PlotRed, thick] table[x=budget, y=diversity, comment chars={\%}, col sep=comma] {data/waterRisk.csv};
\addlegendentry{Diversity};
\addplot[no markers, solid, PlotBlue, thick] table[x=budget, y=hardening, comment chars={\%}, col sep=comma] {data/waterRisk.csv};
\addlegendentry{Hardening};
\addplot[no markers, solid, PlotOrange, thick] table[x=budget, y=combination, comment chars={\%}, col sep=comma] {data/waterRisk.csv};
\addlegendentry{Combination};
\end{axis}
\end{tikzpicture}
\caption{Security risk in the water-distribution network when investing only in redundancy, only in diversity, only in hardening, or in their combination.}
\label{fig:waterRisk}
\end{figure}

\ifExtended
\subsubsection{Water-Distribution Network}
First, we study risks in the water-distribution network.
\fi
Figure~\ref{fig:waterRisk} shows the security risk in the water-distribution network for various budget values invested into the canonical approaches (i.e., redundancy, diversity, or hardening) and their optimal combination.
Again, we note the logarithmic scaling on the vertical axis.
We see that investing in a combination of redundancy, diversity, and hardening results in significantly lower risks than investing in only one of these approaches, thus demonstrating the efficacy and superior performance of a synergistic approach. 

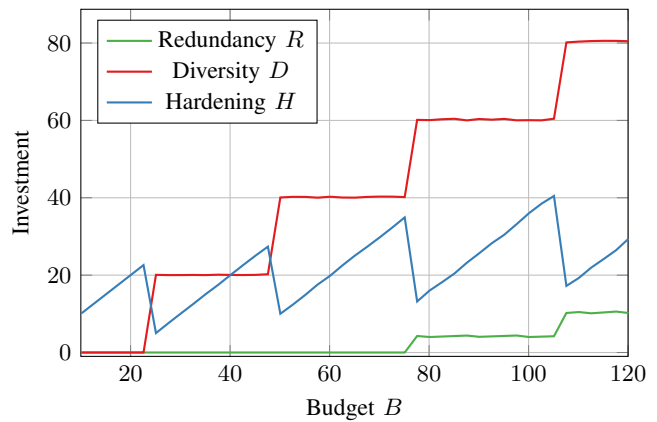
\begin{figure}[h!]
\begin{tikzpicture}[semithick]
\begin{axis}[
  width=\columnwidth,
  height=\figureHeight,
  legend pos=north west,
  xlabel=Budget $B$,
  ylabel=Investment,
  grid=major,
  font=\small,
  xmin=10,
  xmax=120,
  ymin=-1
]
\addplot[no markers, solid, PlotGreen, thick] table[x=budget, y=redundancy, comment chars={\%}, col sep=comma] {data/waterCombination.csv};
\addlegendentry{Redundancy $R$};
\addplot[no markers, solid, PlotRed, thick] table[x=budget, y=diversity, comment chars={\%}, col sep=comma] {data/waterCombination.csv};
\addlegendentry{Diversity $D$};
\addplot[no markers, solid, PlotBlue, thick] table[x=budget, y=hardening, comment chars={\%}, col sep=comma] {data/waterCombination.csv};
\addlegendentry{Hardening $H$};
\end{axis}
\end{tikzpicture}
\caption{Optimal combination of redundancy, diversity, and hardening investments in the water-distribution network.}
\label{fig:waterCombination}
\end{figure}

\ifExtended
\begin{figure}
\centering
\ifExtended
\else
\vspace{-1.5em}
\fi
\ifExtended
\includegraphics[width=0.8\columnwidth]{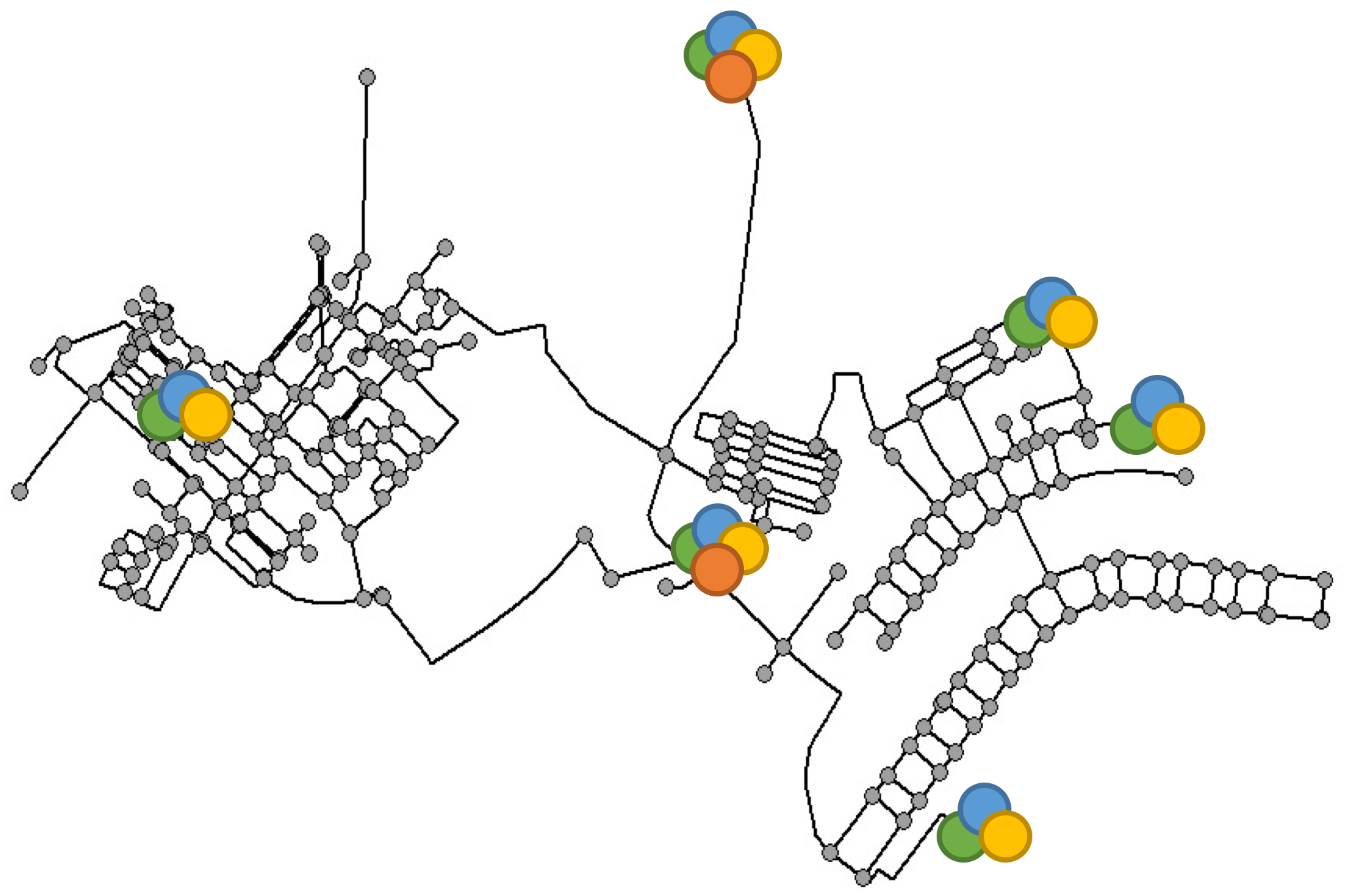}
\else
\includegraphics[width=0.6\columnwidth]{images/deployment.png}
\fi
\ifExtended
\else
\vspace{-0.5em}
\fi
\caption{Optimal deployment with budget $B = 90$.}
\label{fig:deployment}
\end{figure}
\fi

Figure~\ref{fig:waterCombination} shows the optimal combination of redundancy, diversity, and hardening investments in the water-distribution network for various budget values.
In this example, the optimal design is primarily a combination of diversity and hardening.
However, with higher budget values, designers also need to invest in redundancy.
\ifExtended
Note that the design approach also determines the optimal deployment of components.
Figure~\ref{fig:deployment} shows the optimal deployment for budget $B = 90$.
Colored disks represent component instances, different colors corresponding to different implementations.
\fi

\ifExtended
\subsubsection{Transportation Networks}
Second, we consider security risks in the transportation network.
In this case, we restrict our study to diversity and hardening since deploying multiple instances of a traffic light may be infeasible in practice.
Hence, we assume that exactly one instance is deployed for each component.

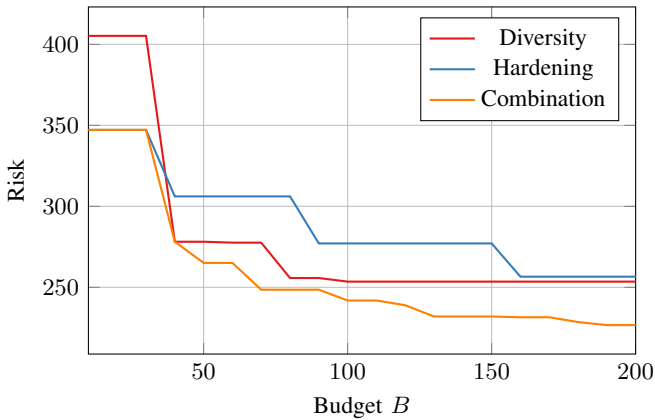
\begin{figure}[h!]
\begin{tikzpicture}[semithick]
\begin{axis}[
  width=\columnwidth,
  height=\figureHeight,
  legend pos=north east,
  xlabel=Budget $B$,
  ylabel=Risk,
  grid=major,
  font=\small,
  xmin=10,
  xmax=200
]
\addplot[no markers, solid, PlotRed, thick] table[x=budget, y=diversity, comment chars={\%}, col sep=comma] {data/transportationRisk.csv};
\addlegendentry{Diversity};
\addplot[no markers, solid, PlotBlue, thick] table[x=budget, y=hardening, comment chars={\%}, col sep=comma] {data/transportationRisk.csv};
\addlegendentry{Hardening};
\addplot[no markers, solid, PlotOrange, thick] table[x=budget, y=combination, comment chars={\%}, col sep=comma] {data/transportationRisk.csv};
\addlegendentry{Combination};
\end{axis}
\end{tikzpicture}
\ifExtended
\else
\vspace{-1.2em}
\fi
\caption{Security risk in the transportation network when investing only in diversity, only in hardening, or in their combination.}
\label{fig:transportationRisk}
\end{figure}

Figure~\ref{fig:transportationRisk} shows the security risk in the transportation network with the canonical approaches and their combinations for various budget values.
The figure shows that---similar to the case of water-distribution networks---the combined approach is clearly superior to canonical approaches.

\begin{figure}[h!]
\begin{tikzpicture}[semithick]
\begin{axis}[
  width=\columnwidth,
  height=\figureHeight,
  legend pos=north west,
  xlabel=Budget $B$,
  ylabel=Investment,
  grid=major,
  font=\small,
  xmin=10,
  xmax=200,
  ymin=-1
]
\addplot[no markers, solid, PlotRed, thick] table[x=budget, y=diversity, comment chars={\%}, col sep=comma] {data/transportationCombination.csv};
\addlegendentry{Diversity $D$};
\addplot[no markers, solid, PlotBlue, thick] table[x=budget, y=hardening, comment chars={\%}, col sep=comma] {data/transportationCombination.csv};
\addlegendentry{Hardening $H$};
\end{axis}
\end{tikzpicture}
\ifExtended
\else
\vspace{-1.2em}
\fi
\caption{Optimal combination of diversity and hardening investments in the transportation network.}
\label{fig:transportationCombination}
\end{figure}
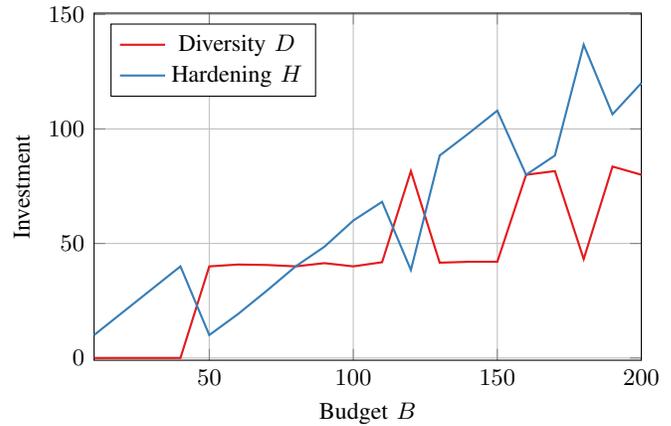

Figure~\ref{fig:transportationCombination} shows the optimal combination of diversity and redundancy in the transportation network for various budget values.
Except for very low values, the optimal combination invests substantial amounts in both diversity and hardening.
\fi

\ifExtended
\subsection{Performance}
\fi

To illustrate the performance of the proposed design algorithm, 
we use the water-distribution network with $R = 10$ and $D = H = 100$.
We find that the meta-heuristic algorithm (Algorithm~\ref{alg:sim_ann}) is very efficient: a single iteration takes less than $6.4 \times 10^{-4}$ seconds (more than 1,500 iterations per second) on an average laptop computer\ifExtended\footnote{MacBook Pro with 2.9 GHz Intel Core i5 processor.}\fi.
To determine the number of iterations that are necessary to find a good solution in practice, we focus on the solution quality (i.e., security risk) as a function of the number of iterations.

\begin{figure}[h!]
\begin{tikzpicture}[semithick]
\begin{axis}[
  width=\columnwidth,
  height=\figureHeight,
  ymode=log,
  xmax=1200,
  xlabel=Iteration,
  ylabel=Risk,
  font=\small,
  grid=major
]
\addplot[no markers, solid, red, thick] table[x=iteration, y=current, comment chars={\%}, col sep=comma] {data/waterSimAnn.csv};
\addlegendentry{Current solution};
\addplot[no markers, solid, blue, thick, dashed] table[x=iteration, y=best, comment chars={\%}, col sep=comma] {data/waterSimAnn.csv};
\addlegendentry{Best solution};
\end{axis}
\end{tikzpicture}
\caption{Security risk in each iteration of one execution of the the meta-heuristic algorithm (Algorithm~\ref{alg:sim_ann}).} 
\label{fig:waterSimAnn}
\end{figure}
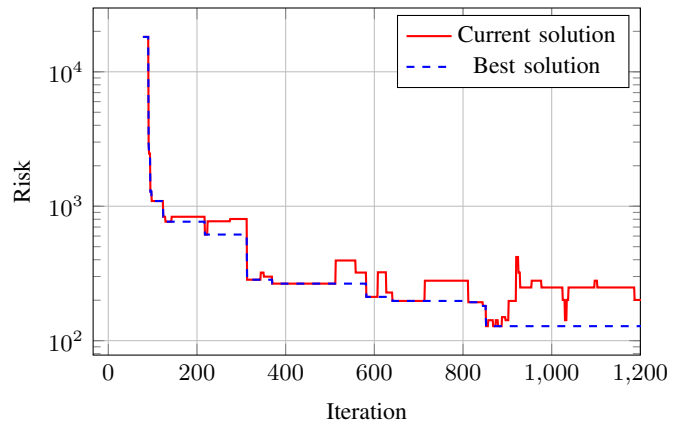

Figure~\ref{fig:waterSimAnn} shows the security risk in each iteration of one particular execution of the meta-heuristic algorithm (Algorithm~\ref{alg:sim_ann}) with the current solution (solid red line) and with the best solution found so far (dashed blue line). 
Please note the logarithmic scaling on the vertical axis.
We have executed the algorithm a number times, but since the results are qualitatively the same, we plot only one particular execution for illustration.
The figure shows that risk decreases rapidly in the first few hundred iterations, but after around 400 iterations, the decrease becomes much slower.
At around one thousand iterations, the risk reached its lowest value, so we omit the remaining iterations from the plot.
In light of this, it is clear that the running time of the meta-heuristic algorithm is very low since it settles in a matter of seconds.

\section{Related Work}
\label{sec:related}

\ifExtended Modern technology trends such as \fi IIoT and cyber-physical systems (CPS) have significantly improved the overall functionality, reliability, observability, and operational efficiency of industrial control systems and critical infrastructure networks \cite{moyne2007emergence,colombo2017industrial}. The integration and connectivity between various system components allow data exchange and information processing to fine tune system processes\ifExtended. However, \fi this integration and connectivity also opens new threat channels in the form of cyber- and cyber-physical attacks, against which these systems need to be secured \cite{sadeghi2015security,koutsoukos2018sure}. Conventional cybersecurity mechanisms are inadequate and thus need to be expanded to incorporate the complexity and physical aspects of such systems \cite{sadeghi2015security,koutsoukos2018sure,cheminod2013review}. 
A detailed overview of the security issues in industrial automation systems that are based on  open communication systems is provided in \cite{dzung2005security}. Similarly, security issues associated with various documented standards in SCADA systems are highlighted in \cite{gao2014scada,cherdantseva2016review}, and it is concluded that such issues cannot be resolved by employing only IT security mechanisms. There are various other studies that mainly highlight the security threats and associated risk assessment in the domain of industrial IoT, for instance \ifExtended\cite{da2014internet,meltzer2015securing,lee2015cyber,jing2014security,suo2012security}\fi.  
All of these studies discuss and point towards a holistic security framework to address the security issues in industrial IoT. In this paper, we provide a framework for synergistic security that combines various security mechanisms to effectively secure such systems.

The water-supply industrial sector can benefit significantly from applying the ideas and technology of industrial Internet \cite{Kartakis2016}. 
\ifExtended
An intelligent urban water-supply management system, which consists of IoT gateways connecting the water assets (for instance, water pumps, valves, and tanks) to the cloud service platform for advanced analytics, significantly improves the operational efficiency, safety, and service availability of the overall system \cite{WaterTestBed,liu2014computational}. There are ongoing efforts to develop efficient remote monitoring systems for pipeline monitoring (such as PIPENET deployed at Boston Water and Sewer Commission  \cite{stoianov2007pipenet,stoianov2008sensor}), water quality monitoring \cite{ali2015network,mckenna2008detecting,storey2011advances}, leak and burst detection \cite{perez2014leak,perelman2016sensor}, and other applications, for instance \cite{yoon2011swats,torbol2013remote,suciu2017unified}. 
\fi
The adoption of new technologies (such as IoT, CPS) and networking devices enhances the monitoring capability, service reliability, and operational efficiency of water distribution systems, but also exposes them to malicious intrusions in the form of cyber- and cyber-physical attacks \ifExtended\cite{taormina2017characterizing,perelman2014network,amin2013cyberCST}\fi. A number of attack scenarios against water distributions systems are specified and demonstrated through simulations in \cite{taormina2017characterizing}. Recently, in \cite{antonioli2018taking}, several attacks on simulated and a real water distribution testbed (WADI \cite{ahmed2017wadi}) are demonstrated through cyber-physical botnets capable of performing adversarial control strategies under CPS constraints. The security breach in the SCADA system of Maroochy Water Services, Australia \cite{slay2007lessons} is a famous incident, which also highlights the need for effective security mechanisms. To effectively address the security challenge in such complex, interconnected, and spatially expanded systems, we need to employ a combination of security mechanisms to protect them against cyber-physical attacks.

\ifExtended
Like other modern infrastructures, traffic networks are complex and are becoming increasingly connected with traffic lights, road sensors, and vehicles exchanging information with each other. This interconnectedness---though useful at many levels---has also increased the attack surface for potential attackers that can significantly disrupt the traffic by taking control of a few network components, such as signal lights or sensors \cite{laszka2016vulnerability,ghena2014green,ghanavati2017analysis}. Recent studies outline the scope of the damage that can be caused by an adversary having an access to the traffic control infrastructure \cite{reilly2016creating}. There are studies demonstrating attacks that can realize non-existent jams and virtual vehicles, tamper with signal schedules \cite{jeske2013floating,grad2009engineers,zetter2014hackers,tufnell2014students}. Considering the impact of successful attacks, it is imperative to systematically understand the existence of vulnerabilities, and design security frameworks to protect traffic infrastructure against such malicious attacks \cite{ernst2017framework,feng2018vulnerability}.
\fi

\ifExtended
\section{Conclusion}
\label{sec:concl}

In this paper, we introduced a framework that considers three canonical approaches--redundancy, diversity, and hardening--for improving 
security and resilience of IIoT systems.
Our goal is to provide theoretical foundations for designing systems 
that combine these canonical approaches.
We showed that the problem of finding an optimal design is computationally hard, which means that practical designs may not be found using exhaustive searches.
Therefore, we introduced an efficient meta-heuristic algorithm, whose running time is polynomial in the size of the problem instance.
To illustrate the practical applicability of our results, we discussed two example application domains, water distribution and transportation systems.
Our numerical evaluation shows that integrating redundancy, diversity, and hardening can lead to reduced security risk at the same cost. 
\fi


\end{document}